\documentclass[a4paper,12pt]{article}
\pdfoutput=1 

\usepackage{myjinst}
\usepackage{color}

\newcommand{\red}{\protect\color{red}}
\newcommand{\blue}{\protect\color{blue}}
\newcommand{\black}{\protect\color{black}}

\renewcommand\blue\red
\renewcommand\red\black

\title{Full Monte-Carlo
description of the Moscow State University
Extensive Air Shower experiment}

\author[a]{Yu.\,A.\,Fomin,}
\author[a]{N.\,N.\,Kalmykov,}
\author[b]{I.\,S.\,Karpikov,}
\author[a]{G.\,V.\,Kulikov,}
\author[b]{M.\,Yu.\,Kuznetsov,}
\author[b]{G.\,I.\,Rubtsov,}
\author[a]{V.\,P.\,Sulakov}
\author[b]{and S.\,V.\,Troitsky}

\affiliation[a]{D.V.~Skobeltsyn Institute of Nuclear Physics, \\
M.V.~Lomonosov Moscow State University, Moscow 119991, Russia}
\affiliation[b]{Institute for Nuclear Research of the Russian Academy of
Sciences,\\
60th October Anniversary prospect 7A, 117312 Moscow, Russia}

\emailAdd{st@ms2.inr.ac.ru}

\abstract{
The
Moscow State University Extensive Air Shower (EAS-MSU) array
studied high-energy
cosmic rays
with primary energies $\sim (1 - 500)$~PeV
in the Northern hemisphere.
The EAS-MSU data are being revisited following recently found indications
to an excess of muonless showers, which may be interpreted as the first
observation of cosmic gamma rays at $\sim 100$~PeV. In this paper, we
present a complete Monte-Carlo model of the surface detector which
results in a good agreement between data and simulations. The model allows
us to study the performance of the detector and will be used
 \red to obtain physical results in further studies. \black}


\arxivnumber{1607.00309} 

\begin{document}
\maketitle
\flushbottom

\section{Introduction}
\label{sec:intro}

The EAS-MSU array \cite{EAS-MSU} was established in late 1950s and
has been upgraded in early 1980s. The array
aimed at investigations of extensive air showers (EAS)
produced by primary particles in the energy range
$(10^{15}-5\times10^{17})$~eV.
The array operated until 1990 and its main results have been published,
notably the discovery of the knee in the cosmic ray spectrum \cite{knee}
by the early version of the installation, results on the primary
\red spectrum \cite{EAS-MSU-spec} and \black chemical composition in the
knee energy region \cite{EAS-MSU-results, EAS-MSU-results1}. A unique
feature of the array was the presence of large-area underground muon
detectors sensitive to muons with energies $\gtrsim 10$~GeV. Quite
recently, an analysis of the data of these detectors, \red whose area and
energy threshold have \black no analogs in modern installations, revealed
an excess of muonless events which may be interpreted as an evidence for
showers initiated by primary photons \cite{EAS-MSU-gamma1, EAS-MSU-gamma2,
EAS-MSU-gamma3}. If confirmed, this observation would mean the discovery
of first cosmic gamma rays above 100~TeV. Therefore, it calls for a
careful reanalysis. Indeed, muonless or muon-poor events may appear as
rare fluctuations of usual, hadron-induced air showers. The estimate of
this background represents a crucial ingredient in a reliable study of
photon candidate events. Previous studies \cite{EAS-MSU-gamma1,
EAS-MSU-gamma2, EAS-MSU-gamma3} used a simplified model of the detector
which, in principle, might underestimate rare fluctuations in the muon
content of hadronic showers. Here, we develop a modern Monte-Carlo
description of the installation. It is based on the air-shower simulation
with CORSIKA \cite{CORSIKA} supplemented by a model of the detector. As is
customary in modern EAS experiments, see e.g.\ Ref.~\cite{Ben}, the result
of a simulation run is recorded in precisely the same format as the real
data. This record is further processed by the usual reconstruction
routine, so that real and simulated events are processed with the same
analysis code. A simulation of this kind was not technically possible at
the time the bulk of the installation's results were obtained. The aim of
this paper is to present the simulation, to estimate the performance of
the installation and to demonstrate a good agreement between real and
simulated data in terms of basic reconstruction parameters, which opens
the way to use the Monte-Carlo description in further studies of the
EAS-MSU data. Results of these physics studies will be reported in further
publications.

The rest of the paper is organized as follows. In Sec.~\ref{sec:array},
the EAS-MSU array is described in detail. Section~\ref{sec:reconstruction}
discusses the event reconstruction procedure, including quality cuts used
in the analysis of both real data and simulated events. In
Sec.~\ref{sec:MC}, we turn to the Monte-Carlo procedure and describe both
the air-shower simulation and modelling of the detector.
Section~\ref{sec:comparison} presents a comparison between simulated and
real events, as well as between thrown and reconstructed parameters. We
briefly conclude in Sec.~\ref{sec:concl}.

\section{The EAS-MSU array}
\label{sec:array}
The installation worked in various configurations. Here, we present
a description of the array relevant for the observation period of 1984 to
1990, which is discussed throughout the paper.

\paragraph{The array of counters.}
The array was located in the M.V.~Lomonosov Moscow State University campus
(geographical coordinates of the array center are 37.54E, 55.70N). It
covered the total area of 0.5~km$^{2}$ and consisted of 76
charged-particle detector stations with multiple Geiger-Mueller
counters in each. The counter data were used to reconstruct the total
number of charged particles, $N_{e}$, of EAS employing the empirical
lateral distribution function and to calculate coordinates of the EAS
axis. To extend the range of measured $N_{e}$, 57 \blue detector stations \black (``vans'')
comprised three types of \blue detectors with \black the Geiger-Mueller counters: \blue one detector with \black 24 counters of
0.0018~m$^{2}$ area, \blue one detector with \black 24 counters of 0.01~m$^{2}$ and \blue three detectors with \black 24 counters
of 0.033~m$^{2}$ \blue each \black (hereafter, small, medium and large \blue detectors and \black counters,
respectively). \red In the very center of the array, there is a special
room where 11 large, 10 medium and 10 small detectors are located. In
reconstruction, they group into 4 independent detector stations.
\black Nineteen other \blue detector stations \black (``boxes'') contained \blue 2 large detectors
\black only and were located in the central part of the array.
The total number of Geiger-Mueller counters was about
10000 with the total collecting area of 250~m$^{2}$. To measure the
density of muons with energies above $\sim$10~GeV, groups of similar
Geiger-Mueller counters, each with the area of 0.033~m$^{2}$, were located
at the depth of 40 meters of water equivalent underground. One of the muon
detectors was located in the center of the array and consisted of 1104
counters with the total area of 36.4~m$^{2}$. The other three were located
at distances of 220~m, 300~m and 320~m from the center of the array; each
of them contained 552 similar counters. \red In the present and subsequent
works we consider only the central muon detection system. \black

\paragraph{The scintillator trigger systems.} The EAS-MSU array included
two independent scintillator trigger systems, based on detector stations
with 5-cm thick plastic scintillator. They were used to determine the
arrival direction of a shower. The first one, the central trigger system,
was intended primarily for the selection of showers with particle number
$N_{e} \lesssim 2 \times10^{7}$. It was located in the central part of the
array and consisted of 7 scintillation counters: one with the area of
4~m$^{2}$ in the array center and 6 of the area of 0.5~m$^{2}$ each,
located at distances of about 60~m from the center. The condition to
trigger is a simultaneous, within the time gate of 500~ns, response of the
central scintillation counter and at least two other counters, so that
they form a triangle and not a straight line, to be able to estimate the
arrival direction of EAS. To reduce the trigger rate, the
central trigger system included the additional criterion of the express
analysis: firing of 56 of 264 Geiger-Mueller counters of the area of
0.033~m$^{2}$ located in a special room at the center of the array. If the
number of triggered counters is less than 56, which means that the density
at the array center does not exceed about 6 particles per square meter,
the event was not registered.

The second, peripheral, trigger system was designed for the efficient use
of the entire area of the array for registration of EAS with particle
number $N_{e} \gtrsim 2 \times10^{7}$. It consisted of 22 scintillation
counters of the area of 0.5~m$^{2}$ each, combined into 13 tetragons with
sides of (150--200)~m.
\red
Each of these scintillation counters was located in the one of the "van"
detector stations. \black The criterion to trigger was a simultaneous,
within the time gate of 6~$\mu$s, response in at least one tetragon of
scintillation counters. Similarly to the central one, the peripheral
trigger system included the express analysis: it was required that at
least four \red detector stations of the peripheral trigger system have at
least 4 of 72 large Geiger-Mueller counters fired. \black which
corresponds to the density higher than $\sim 1.7$ particles per square
meter.

\begin{figure}
\begin{minipage}[h]{0.48\linewidth}
\center{\includegraphics[width=1\linewidth]{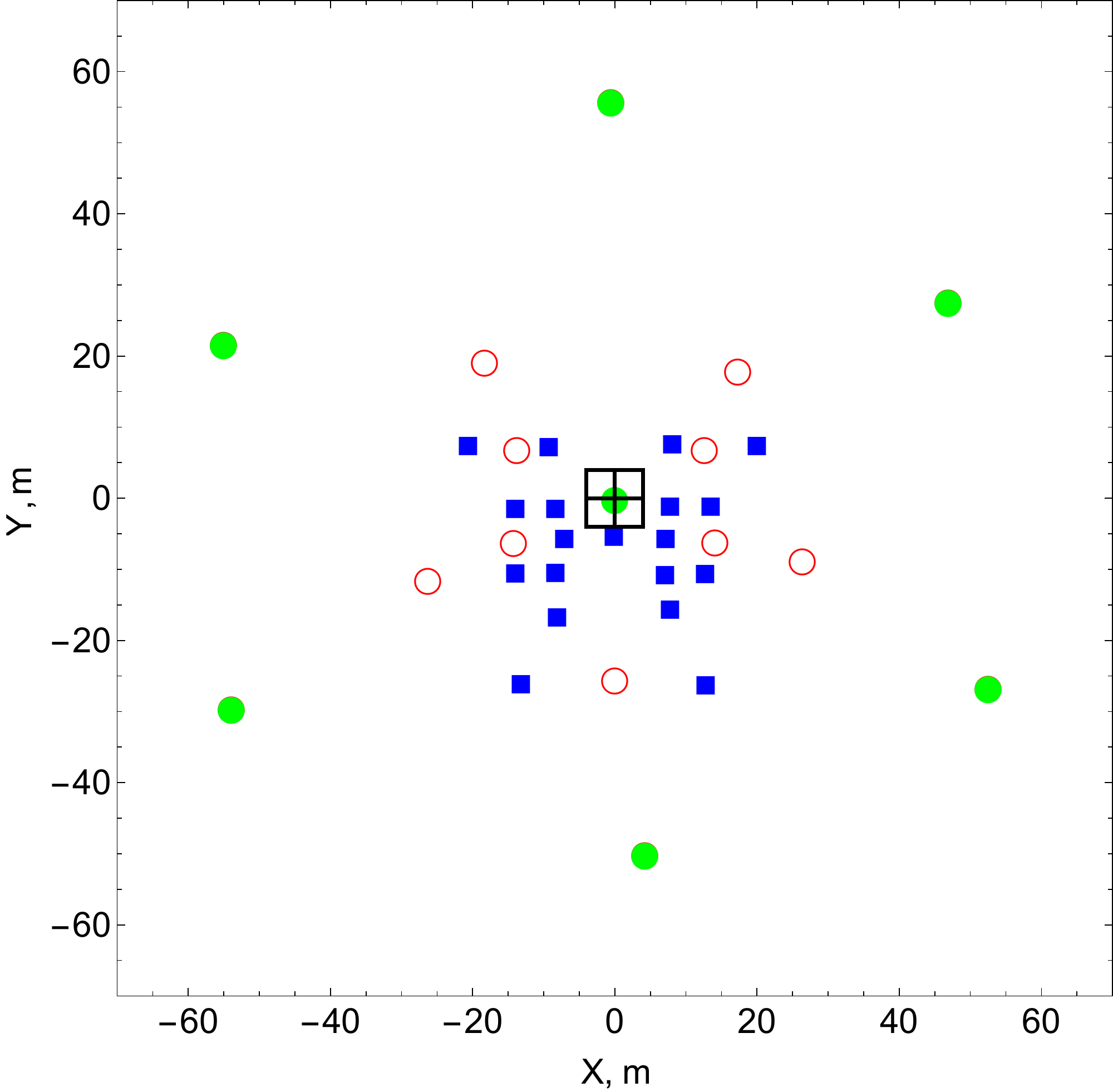} \\
(a)}
\end{minipage}
\hfill
\begin{minipage}[h]{0.48\linewidth}
\center{\includegraphics[width=1\linewidth]{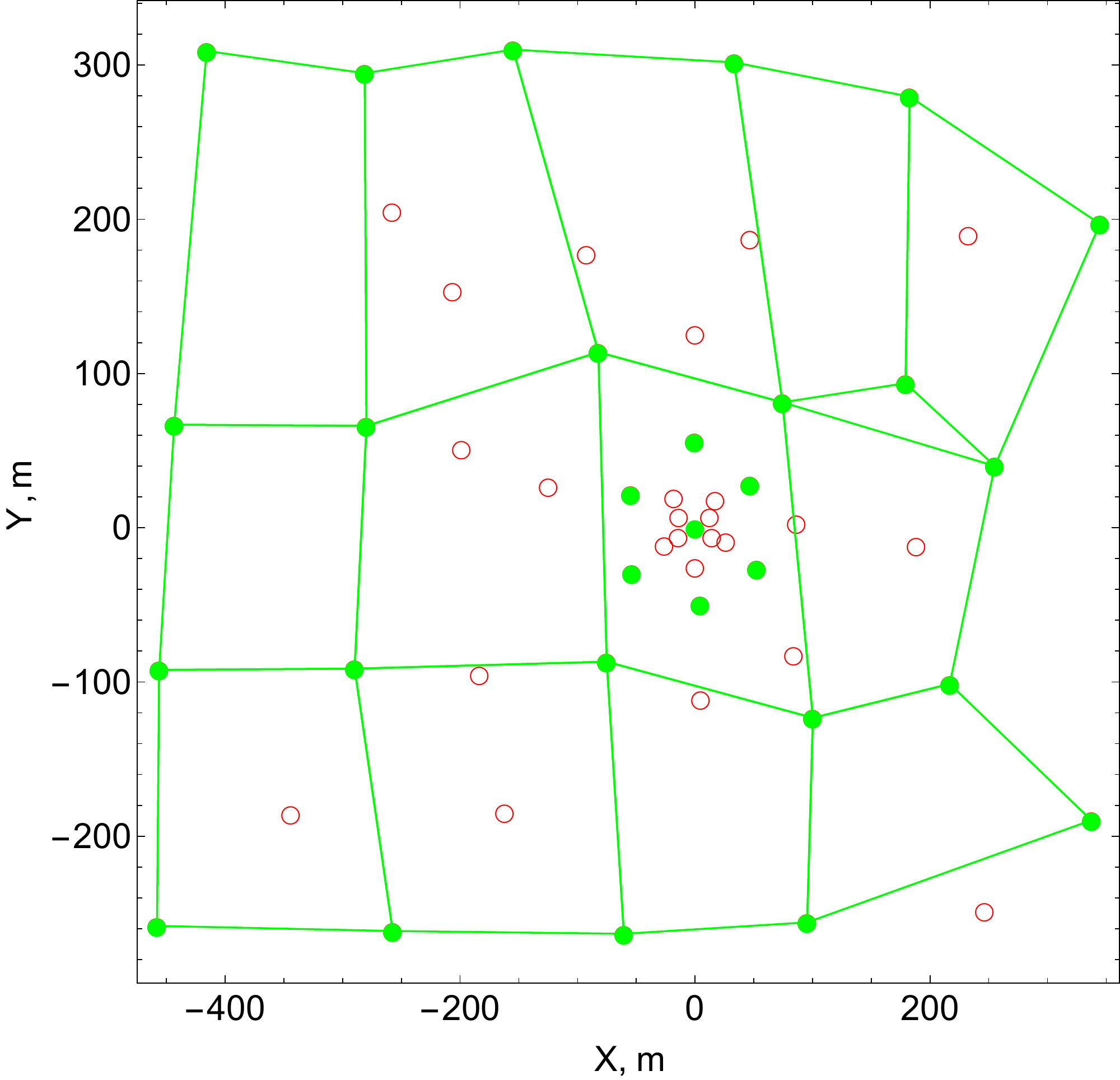} \\
(b)}
\end{minipage}
\caption{The EAS-MSU array setup: (a)~the central part of the array;
(b)~the entire array. Four empty black squares in the center represent the
special \red central \black room.
Green circles denote ``vans'' with scintillation
counters, empty red circles denote ``vans'' without scintillation counters,
blue squares denote ``boxes''. Lines represent the tetragons of the
peripheral trigger system. See the text for more details.}
\label{fig:geometry}
\end{figure}

The geometry of the array is shown in Fig.~\ref{fig:geometry}.

\section{The event reconstruction procedure and selection cuts}
\label{sec:reconstruction}

\paragraph{Reconstruction.} The procedure for \red the \black determination
of \red the \black parameters of a EAS consists of several stages. First, EAS arrival
angles (the zenith angle, $\theta$, and the azimuth angle, $\phi$) are
calculated analytically from the response times of three scintillation
detectors colocated with \red the \black counters which recorded the highest density of
charged particles \red \cite{book}\black. Further, the found angles \red
are used \black as the first approximation for the arrival direction\red.
The \black readings of 10 to 15, depending on the size of the shower,
charged-particle stations with the highest density,
\red together with the angles, allow to determine \black
the EAS axis position $(X,Y)$ in the plane of the array,
the total number of particles $N_{e}$ and the EAS age parameter $S$ \red
in the first approximation. To this end, \black the method
of least squares \red is used for \black the lateral distribution function
described below. Then, by making use of time delays of all triggered
scintillation counters and of the axis position, the EAS arrival direction
is recalculated by the method of the maximum likelihood function  \red
\cite{book}\black. Further, using these new $\theta$ and $\phi$, the EAS
position, $N_{e}$ and $S$ are recalculated. These iterations continue
until the process converges\red\footnote{\red The process is considered as
converged when the difference between the arrival directions calculated
in two subsequent iterations does not exceed 0.005 radian.}\black.

\begin{figure}
\center{\includegraphics[width=1\linewidth]{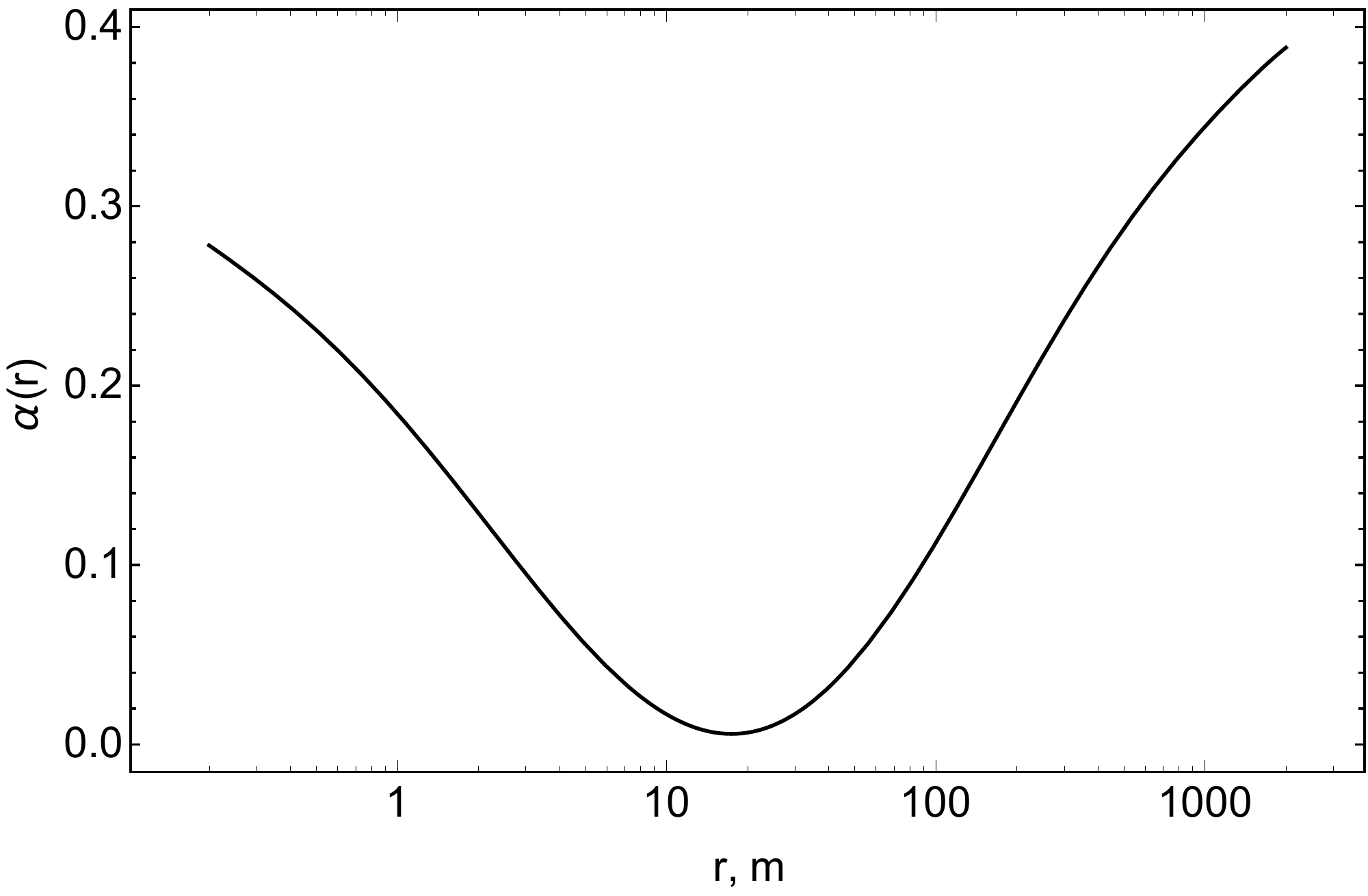}%
}
\caption{%
The correction $\alpha(\red r \black)$ to
the age parameter as a function of the
distance from the shower axis $\red r \black$.
}
\label{fig:LDFfuncs}
\end{figure}

The key ingredient of the reconstruction procedure is the lateral
distribution function (LDF). It was obtained experimentally
\cite{EAS-MSU:LDF} by the analysis of showers with $\theta<30^\circ$.
These
showers were divided into 19 groups according to the total number of
particles, starting with $\log_{10}N_{e}=4.6$ and with 0.2~dex step,
and, for each group, the mean LDF of charged particles was constructed.
Then, the empirical LDF was determined by the following procedure. The
Nishimura--Kamata--Greisen (NKG) \cite{NKG1, NKG2} function was taken as
the
first
approximation. Then, keeping in mind that
the NKG function is relevant for electrons and positrons while the
installation detects other particles as well,  the concept of the local
age parameter, $S_{\rm local}$, was introduced. The correction to the NKG
function was parametrized by the dependence of this $S_{\rm local}$ on the
distance $\red r \black$ to the shower axis, so that the effective $S_{\rm
local}(\red r \black)$ replaces the shower age in the NKG formula. The mean
experimental LDF is flatter than the original NKG function both at
distances $\red r \black>30$~m and $\red r \black<15$~m. Therefore, the
charged-particle LDF we use is
\begin{equation}
\rho(S,\red r \black)=
N_{e}C(S)(\red r \black/R_{0})^{(S+\alpha(\red r \black)-2)}\cdot(\red r \black/R_{0}+1)^{(S+\alpha(\red r \black)-4.5)},
\end{equation}
where $\rho$ is the particle density,
$R_{0}\approx 80$~m is the Moliere radius,
$S$ is the age parameter of the EAS, which corresponds to the NKG age
determined at 15~m$\, \red \lesssim r \lesssim \black\,$30~m, $C(S)$ is
the normalization coefficient, which is calculated numerically, and
$\alpha(\red r \black)$ is the correction to $S$ determined empirically
and shown in Fig.~\ref{fig:LDFfuncs}.

\paragraph{Event selection.} In order to study well-reconstructed showers
with primary energies exceeding $E \sim 10^{17}$~eV, we apply the following
selection cuts:

1. The reconstruction procedure converges and allows to determine the
shower parameters.

2. The age parameter of the EAS is $0.3<S<1.8$.

3. The zenith angle is $\theta<30^\circ$.

4. The distance between the array center and the shower axis is
$R<240$~m.

5. The shower size is $N_{e}>2\times10^7$ (showers with these $N_{e}$
are recorded by the peripheral selection system mostly).

\section{Monte-Carlo simulations}
\label{sec:MC}
The full Monte-Carlo (MC) simulation of the EAS-MSU data includes the
following steps\red, described in more detail below\black:

(i)~simulation of a library of air showers with random arrival directions;

(ii)~generation of random locations of the shower axis;

(iii)~simulation of the response of the detector stations and recording
the simulated event in the format used for real data;

(iv)~reconstruction of the event parameters with the standard procedure
described in Sec.~\ref{sec:reconstruction}.

\paragraph{Air-shower simulations.}
For the first step, we use
the CORSIKA 7.4001 \cite{CORSIKA} simulation
package. For the standard simulated event set, we choose the
QGSJET-II-04 \cite{QGSJET-II-04} as the
high-energy hadronic interaction model and FLUKA2011.2c
\cite{FLUKA1} as the low-energy hadronic interaction
model. The EGS4 \cite{EGS4} electromagnetic model is used for
electromagnetic processes.
\red
As this combination of modern interaction models gives a good description
of the installation, we did not study the effect of change in the models.
\black

At this step, a shower library is produced containing, presently, \red
1370 \black artificial showers (\red 852 \black induced by primary protons
and \red 518 \black induced by primary iron nuclei). Each shower in the
library may be used multiple times at the step (ii). The primary energies
of the showers in the library follow the $E^{-1}$ differential spectrum
with $10^{\red 7.5}$~GeV$<E<10^{\red 8.75}$~GeV. These EAS are simulated
with zenith angles in the range between $0^0$ and $\red 35^0\black$
assuming an isotropic distribution of arrival directions in the celestial
sphere. The showers are simulated without thinning. The lower energy
thresholds are fixed for hadrons (excluding $\pi^{0}$) and muons as
50~MeV; for photons, $e^{+}$, $e^{-}$ and $\pi^{0}$ as 250~keV. The
standard geomagnetic field for the array location is assumed,
$B_{x}=16.5~\mu$T and $B_{z}=49.6~\mu$T. For definiteness, we fix the
model of the atmosphere corresponding to Central Europe for October 14,
1993 (the model number 7 in CORSIKA notations).

\paragraph{Simulation of the detector.}
To further process the CORSIKA showers, we use a
model of the EAS-MSU facility implemented as a C$++$ code.
The showers from the library are randomly selected in such a way that the
resulting differential
spectrum $dN/dE \sim E^{-3.1}$ is obtained. This spectrum was taken as an
approximate fit of combined data presented in Ref.~\cite{Agashe:2014kda},
not taking into account the ``second knee'' feature. For each of the
selected showers, we generate a random position of the center on the ground
in the array area (more precisely, in the rectangle
\red $|X|<280$~m, $|Y|<280$~m, \black
containing all registrations points plus $50$~m in each direction). We
consider three model primary compositions: pure proton, pure iron and a
two-component p/Fe mixture with a ratio of components fixed from
the data/MC comparison, see below.
\red We also tested the original spectrum measured by
EAS-MSU~\cite{EAS-MSU-spec} and approximated as $dN/dE \sim E^{-3.04}$, as
well as the proton/iron mixtures describing the results of other
experiments, namely
KASCADE--Grande \cite{KASCADE-comp} (59\% iron, 41\% proton) and Tunka-133
\cite{Tunka-comp} (51\% iron, 49\% proton for Tunka-133). The impact of
these variations on the behavior of the observables ($R$, $\theta$,
$N_{e}$) was found negligible.
\black

Given the coordinates and momenta of all particles at the ground level
provided by CORSIKA, we test geometrically which of these
particles hit a detector station within the facility. For the
registration points laying at various heights, we
assume that every particle moves with a given velocity and without any
interactions forth/back in the direction it has at the ground
level. We neglect the interactions of particles with materials covering
registration points (typically it is a few millimetres of metal plus a few
centimetres of wood), implying it is a part of the atmosphere.
\red
We also simulate muon detectors located in a deep
underground chamber. For vertically falling particles the shielding of
this array amounts to 40~m of water equivalent. Simulating these detectors
we select only muons ($\mu^{+}$ and $\mu^{-}$; the electromagnetic
contamination is far below one per cent level) whose
continuous-slowing-down approximation (CSDA) range in water exceeds its
path into shielding material. In turn the CSDA range for muon in water is
given by \black
$$(l_{\rm stop}/\mbox{cm}) \simeq 229.69 + 412.14 \,
(E_{\mu}/\mbox{GeV}) - 0.71819 \, (E_{\mu}/\mbox{GeV})^2,$$ where $E_{\mu}$
is the muon energy and $l_{\rm stop}$ is the distance at which the
particle loses all its kinetic energy. The approximation is a fit to the
data of Ref.~\cite{Groom:2001kq}.
\red
For vertical muons the threshold energy yields approximately 10~GeV.
\black
The data/MC comparison for the muon
counters will be studied elsewhere.

In our model, the Geiger--Mueller detector response is simulated as follows.
During the onset of the shower, each counter can be activated only once, due to its
long dead time ($40~\mu$s). Thus the measured value is the number
of activated counters in a \blue detector. \black The outer registration points, ``vans'',
can measure charged-particle density $\rho_{e}$ in limits from $\rho_{e}
\approx 0.42$~m$^{-2}$ to $\rho_{e} \approx 1750$~m$^{-2}$.
The same limits for the inner registration points, ``boxes'', are from
$\rho_{e} \approx 0.64$~m$^{-2}$ to $\rho_{e} \approx 117$~m$^{-2}$.
The counter itself is
a tube of glass with $1$~mm thick wall and $0.1$~mm germanium
spraying from inside. We assume that a  counter is activated for every
charged particle hitting the tube. For photons with energy $E_{\gamma}
\gtrsim 1$~keV, we use a formula for the interaction probability
$$P_{\rm int} = 0.011 - 0.058 / \ln (E_\gamma/\mbox{keV})
+ (E_\gamma/\mbox{keV})^{-0.695}.$$
We obtain it as a numerical approximation of
the data from Ref.~\cite{Agashe:2014kda}, from where we take the total
cross section of photon-induced processes in silicon and germanium with
charged particles in the final state. A subtlety is that both the cross
section and the interaction probability grow with decrease of the photon
energy, making the resulting signal sensitive to low-energy photons. In
order to estimate the impact of the soft photons on the $N_{e}$
reconstruction, we have performed simulations with the 50~keV cutoff for
gamma rays (which is the lowest value allowed by CORSIKA). We obtained
that the contribution of photons to Geiger-Mueller counters activation is
about 10\% of that of charged particles, while in the case of 250~keV
gamma-ray cutoff it is about 7\%.

In the EAS-MSU experiment, each scintillator measured only the time of its
activation. Thresholds for scintillator activation, averaged over
the zenith angle, are $1/3$ and $1$ minimal ionizing particles, for
peripheral scintillators and the central scintillator, respectively. We use
the response functions presented in Ref.~\cite{Sakaki2001}. For photons,
we
make an approximation, randomly choosing only $1/5$ of them and amplifying
the response by the factor of 5. This is justified because of the smooth,
monotonic character of the response function.

The luminescence time of scintillators is $\sim 1~\mu$s (the signal was
integrated over $5~\mu$s), this leads to the accumulation of particles
during the full time of the shower development. For each of the
central-system scintillators, we calculate the time of its activation
relative to the time of the central scintillator activation with 5~ns
resolution. For a peripheral system scintillator, we calculate the time of
its activation relative to the time of the first activated peripheral
scintillator. According to the EAS-MSU native data format, the result is
recorded as a rough part, with 100~ns resolution, and a precise part, with
3.8~ns resolution.

After processing all particles in the shower, the resulting detector
signals are written to a file of the same structure as the experimental
data files. This file has the size of 1024~kb per event. Each
box is encoded in 1~byte which contains a number of
activated counters in this box. The timing of each central system
scintillator is encoded in 1~byte, while the timing of a peripheral-system
scintillator is encoded in 2~bytes, one for the rough part and another for
the precise part. Other bytes in the data file either are empty or contain
technical information (date, time, type of trigger activated etc.)

The simulation of the experiment does not incorporate the daily
\red calibration \black information, which includes the following details.
In the real facility, some detectors were inactive from one day to
another. In our simulation, we neglect this fact because the fraction of
these detectors is not large (about a few percent) at any day. For further
data processing, e.g.\ to estimate primary fluxes, this should be included
in the exposure, so that the effective area of the array is used as a
function of date. We also neglect changes made in cable optical lengths
for the central scintillator system and in time resolution of the precise
part of the peripheral scintillator system timing. Both of these
variations are of order of a few percent and affect the timing record. We
approximate these values by its averages.

\section{Comparison of data with the simulation}
\label{sec:comparison}
An important part of the EAS-MSU array Monte-Carlo simulation is the
comparison of data and MC. This allows to
verify the precision of the simulated event set in its representation of
the real data set and to demonstrate the reliability of event-selection
procedures.  Here, we present results of this comparison for reconstructed
$N_{e}>2 \times 10^{7}$. After applying all quality cuts, the real data
sample contains 922 events, while the MC sample contains \red 4468 \black
proton-induced and \red 1093 \black iron-induced showers. They were
generated from \red 852 \black and \red 518 \black independent CORSIKA
showers, respectively. When stated explicitly, the data/MC comparison is
performed separately for proton and iron primaries. However, as we will
discuss in detail below, the data allow for determination of the best-fit
primary composition experimentally, from the distribution of the age
parameter $S$, assuming the two-component proton-iron mixture. This
best-fit composition (43\% protons and 57\% iron, see item 3 below for its
determination) is used for most of the tests throughout the paper.

\red It is instructive to trace the effect of various cuts on the number
of events. This is done in Table~\ref{tab:cuts}.
\begin{table}[htbp]
\centering
\caption{\label{tab:cuts} \red
Number of independent and resampled CORSIKA showers and the effect of
various cuts on the number of events in Monte-Carlo and in real data. Note
that we are interested in $N_{e}>2\times 10^{7}$ and simulated showers
with $E>10^{16.5}$~eV only, while the installation recorded also events
with much lower energies, which explains a large amount of data events
not passed the $N_{e}$ cut. The geometric area where the artificial
showers are thrown 1.3 times exceeds the area where they are selected.} \smallskip
\begin{tabular}{|c|ccc|}
\hline
&  MC, p & MC, Fe & data \\
\hline
CORSIKA EAS & 852 & 518 & -- \\
sampled EAS & 63060 & 21481 & -- \\
triggered EAS  & 58927 & 21073 & 892321 \\
reconstructed EAS,$0.3<S<1.8$ & 58814 & 21052 & 843086 \\
$R<240$~m & 37987 & 13574 & 702238 \\
$\theta<30^{\circ}$ & 29283 & 10249 & 546493 \\
$N_{e}>2\times 10^{7}$ & 4468 & 1093 & 922 \\
\hline
all cuts & 4468 & 1093 & 922 \\
\hline
\end{tabular}
\end{table}
\black

\begin{figure}
\begin{minipage}[h]{0.48\linewidth}
\center{
\includegraphics[width=1\linewidth]{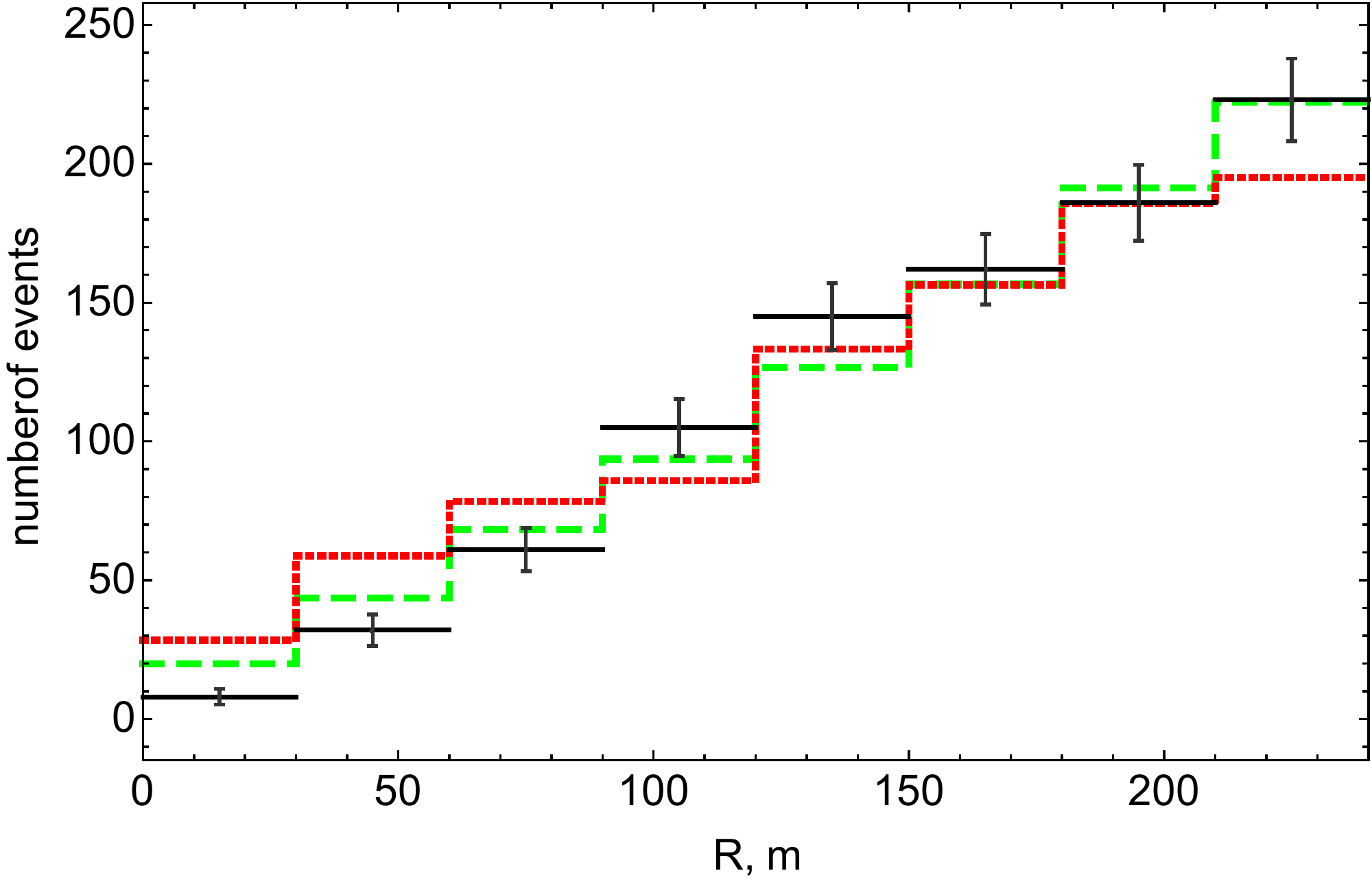} \\
\red (a) \black}
\end{minipage}
\hfill
\begin{minipage}[h]{0.48\linewidth}
\center{\includegraphics[width=1\linewidth]{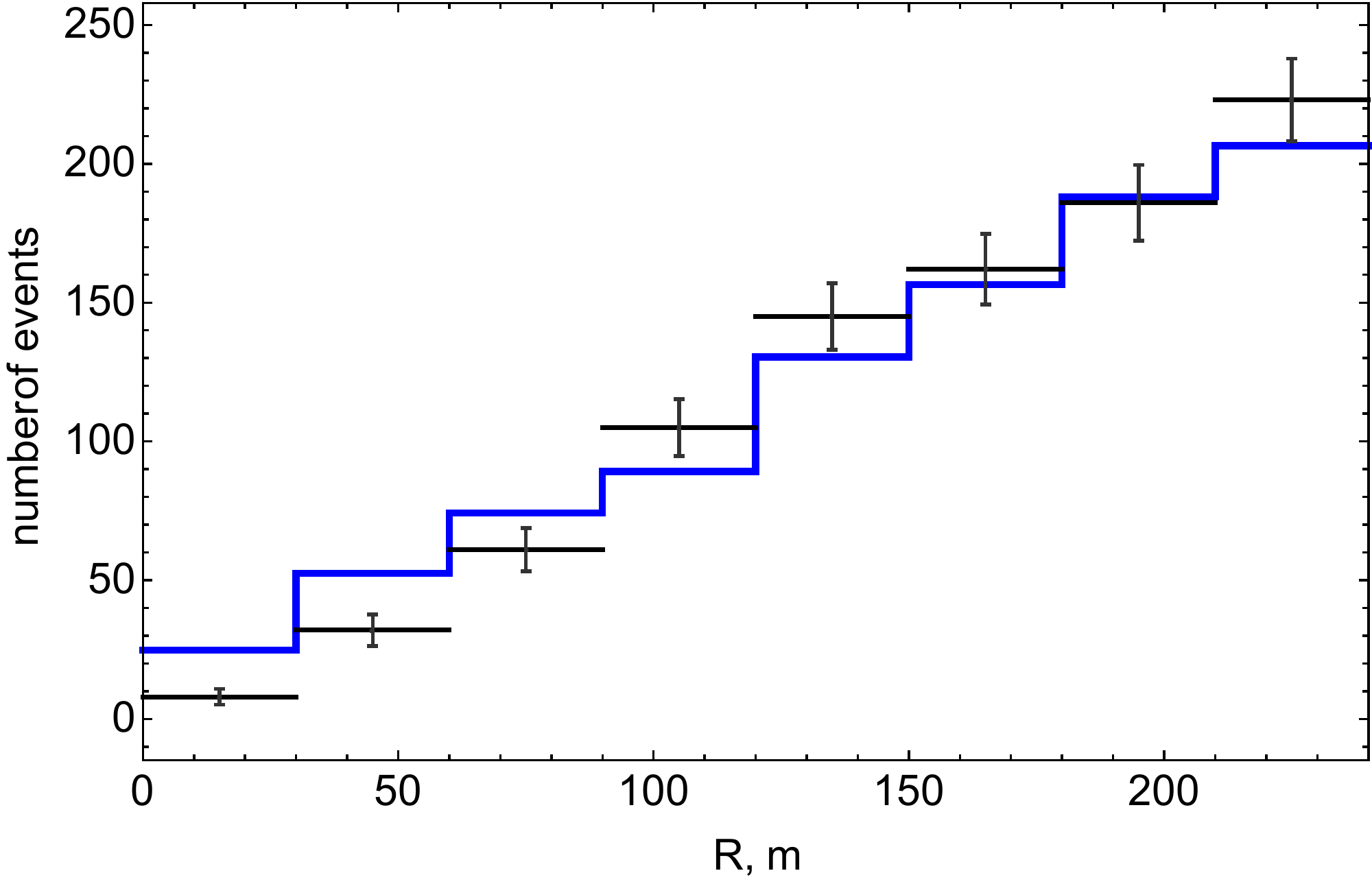}%
\\
(b)
}
\end{minipage}
\caption{Data versus MC comparison of the distribution in $R$.
Points with error bars: data.
\red
(a):~green dashed hystogram: MC (protons), red dotted hystogram: MC
(iron); (b):~blue hystogram: MC (the best-fit composition). }
\black
\label{fig:R-comparison}
\end{figure}

\begin{figure}
\begin{minipage}[h]{0.48\linewidth}
\center{
\includegraphics[width=1\linewidth]{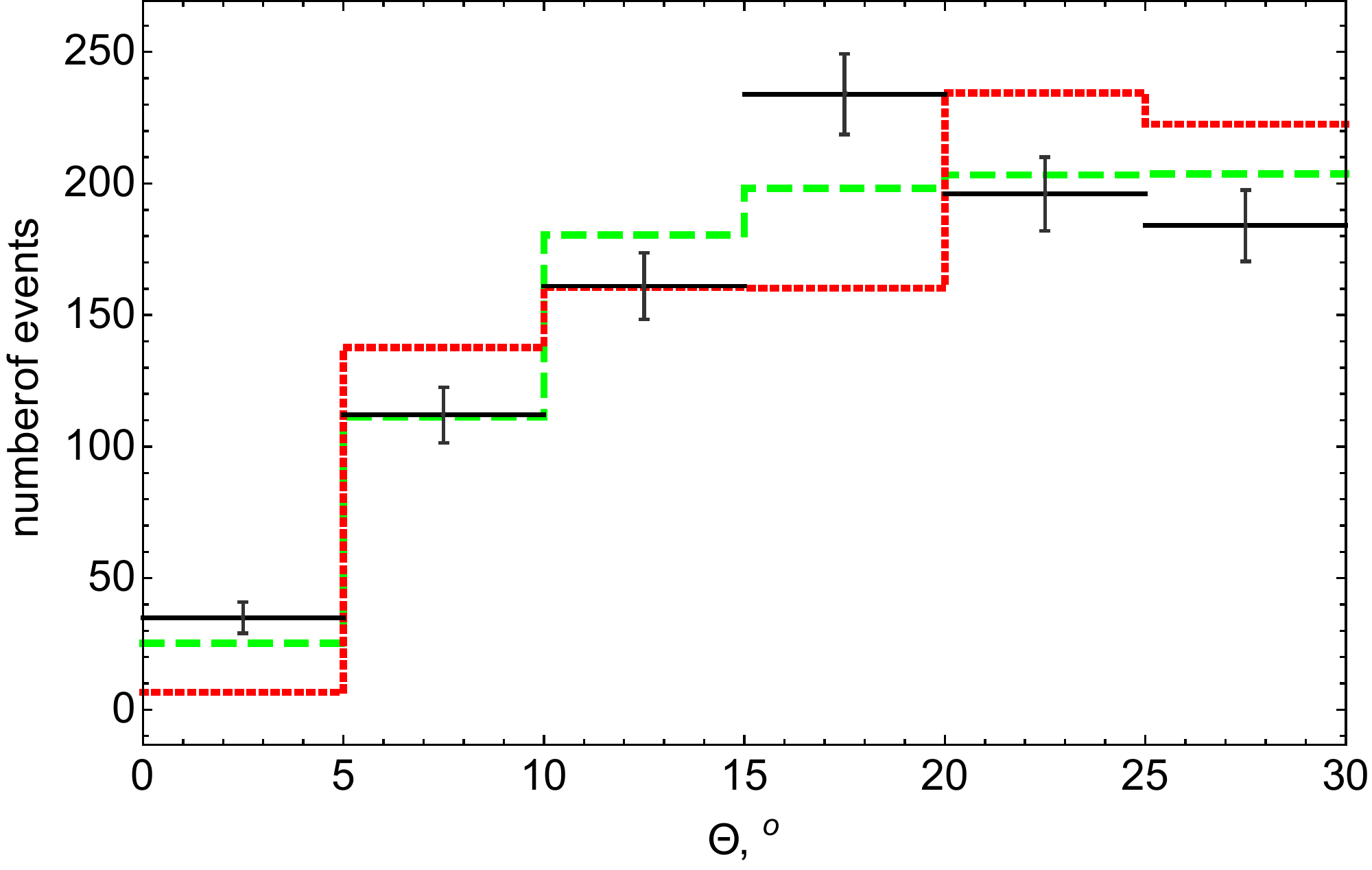} \\
(a)}
\end{minipage}
\hfill
\begin{minipage}[h]{0.48\linewidth}
\center{\includegraphics[width=1\linewidth]{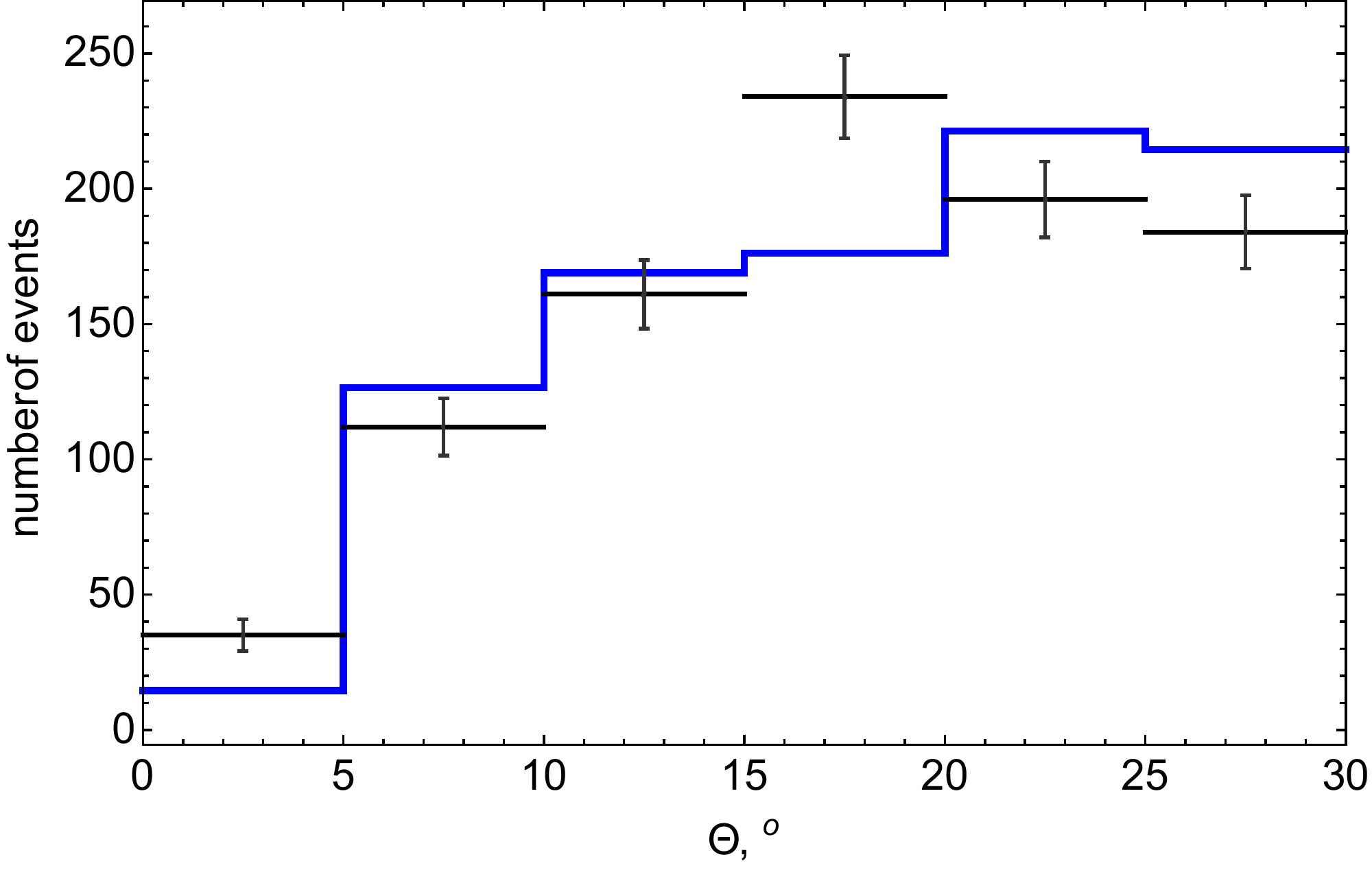}%
\\
(b)
}
\end{minipage}
\caption{Data versus MC comparison of the distribution in $\theta$.
Points with error bars: data.
(a):~green dashed hystogram: MC (protons), red dotted hystogram: MC (iron);
(b):~blue hystogram: MC (the best-fit composition).}
\label{fig:theta-comparison}
\end{figure}

\begin{figure}
\center{
\includegraphics[width=1\linewidth]{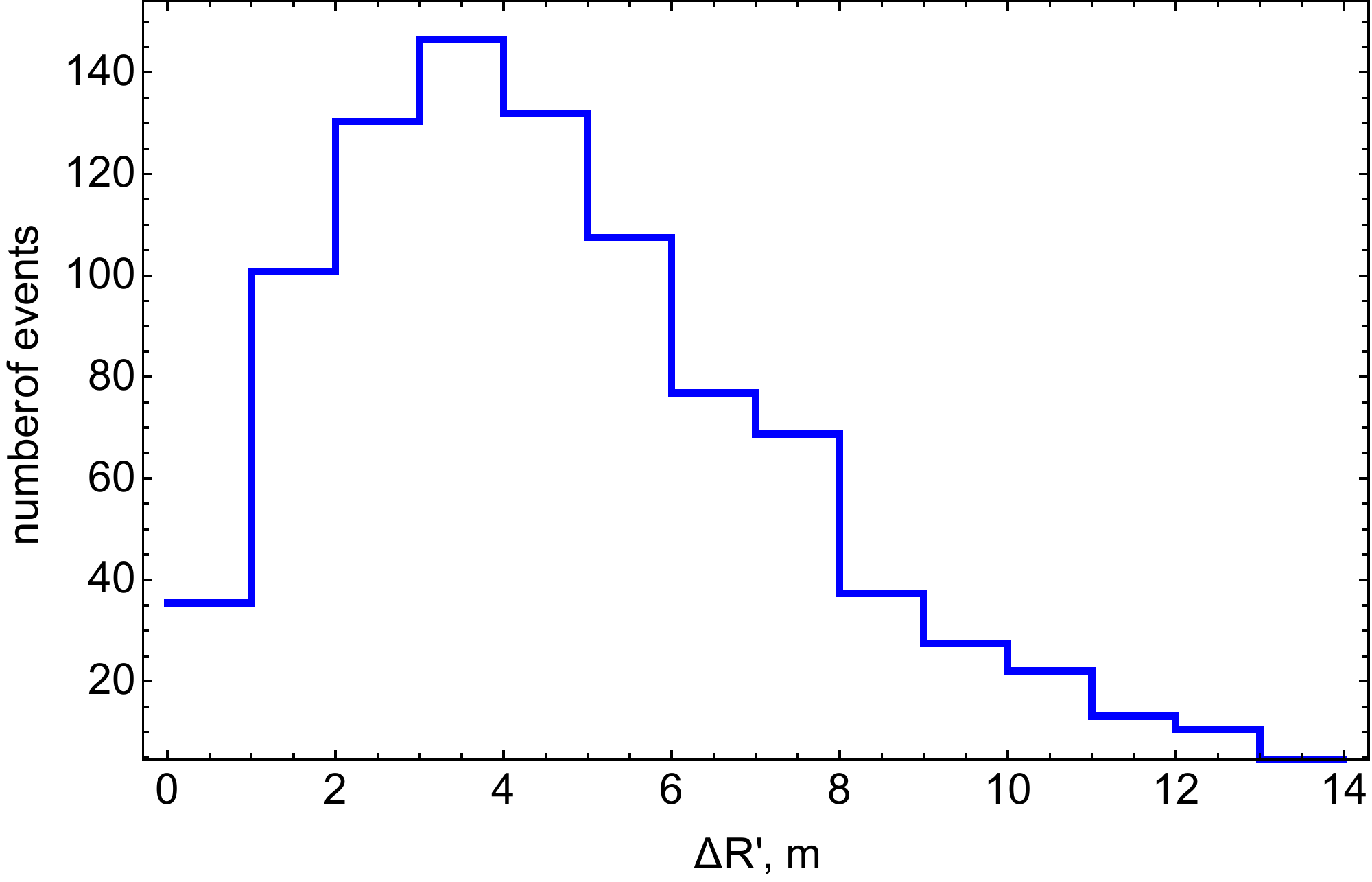}%
}
\caption{Distribution of distances between thrown and reconstructed MC
position of the point where the shower axis crosses the array plane, in
meters (the best-fit composition).
}
\label{fig:axis-precision}
\end{figure}

\begin{figure}
\center{
\includegraphics[width=1\linewidth]{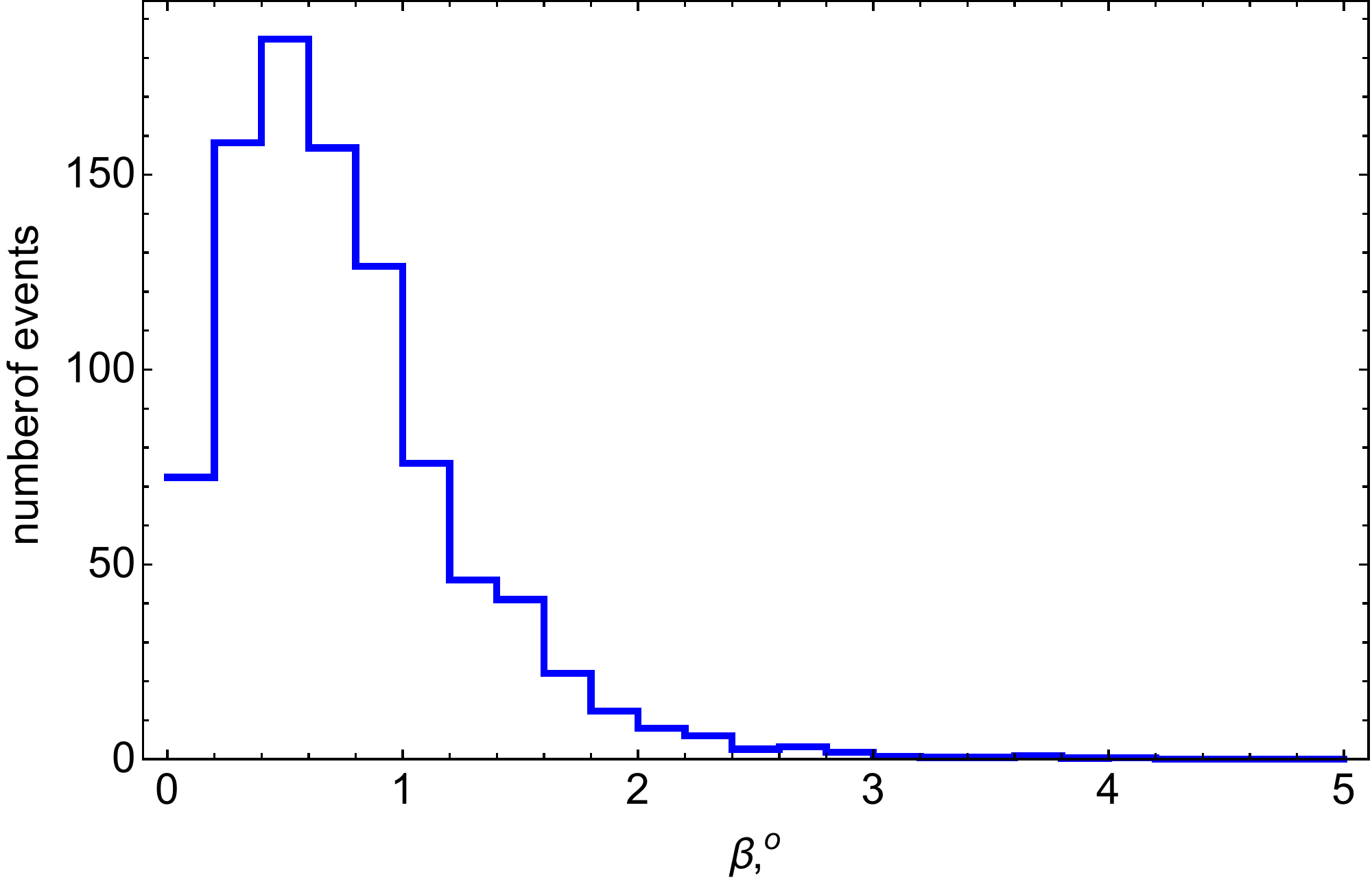}%
}
\caption{Distribution of the angular separation between thrown
and reconstructed MC arrival direction, in degrees (the best-fit
composition).
}
\label{fig:direction-precision}
\end{figure}

\paragraph{1. Geometry and the primary arrival direction.}
The key observables of an air shower are the arrival direction and the
core position; all other recorded parameters are very sensitive to
the geometry.
 Figures~\ref{fig:R-comparison} and
\ref{fig:theta-comparison}
present comparisons between the distributions of the real and MC events in
the distance between the axis and the array center $R$ and the zenith angle
$\theta$,
respectively \red (all histograms in the paper are normalized to the number
of events in the real data sample after all cuts, that is 922) \black.
Comparison between the known trajectory of the thrown MC primary particle
and its reconstructed parameters allows one to determine the accuracy with
which the position of the shower axis and the arrival direction are
determined, see Figs.~\ref{fig:axis-precision} and
\ref{fig:direction-precision}, respectively.
The estimated reconstruction accuracy is presented in
Table~\ref{tab:accuracy}.
\begin{table}[htbp]
\centering
\caption{\label{tab:accuracy} Accuracy of the reconstruction for $N_{e}> 2
\times 10^{7}$ determined from the Monte-Carlo simulations (the best-fit
composition). 68\% of the events are reconstructed with the precision not
worse than the quoted values.} \smallskip
\begin{tabular}{|c|c|}
\hline
axis position, m &  5.7 \\
arrival direction, degree & 1.1 \\
$\Delta N_{e}/N_{e}$ &  0.165 \\
$\Delta E/E$ & 0.41 \\
\hline
\end{tabular}
\end{table}

\begin{figure}
\begin{minipage}[h]{0.48\linewidth}
\center{
\includegraphics[width=1\linewidth]{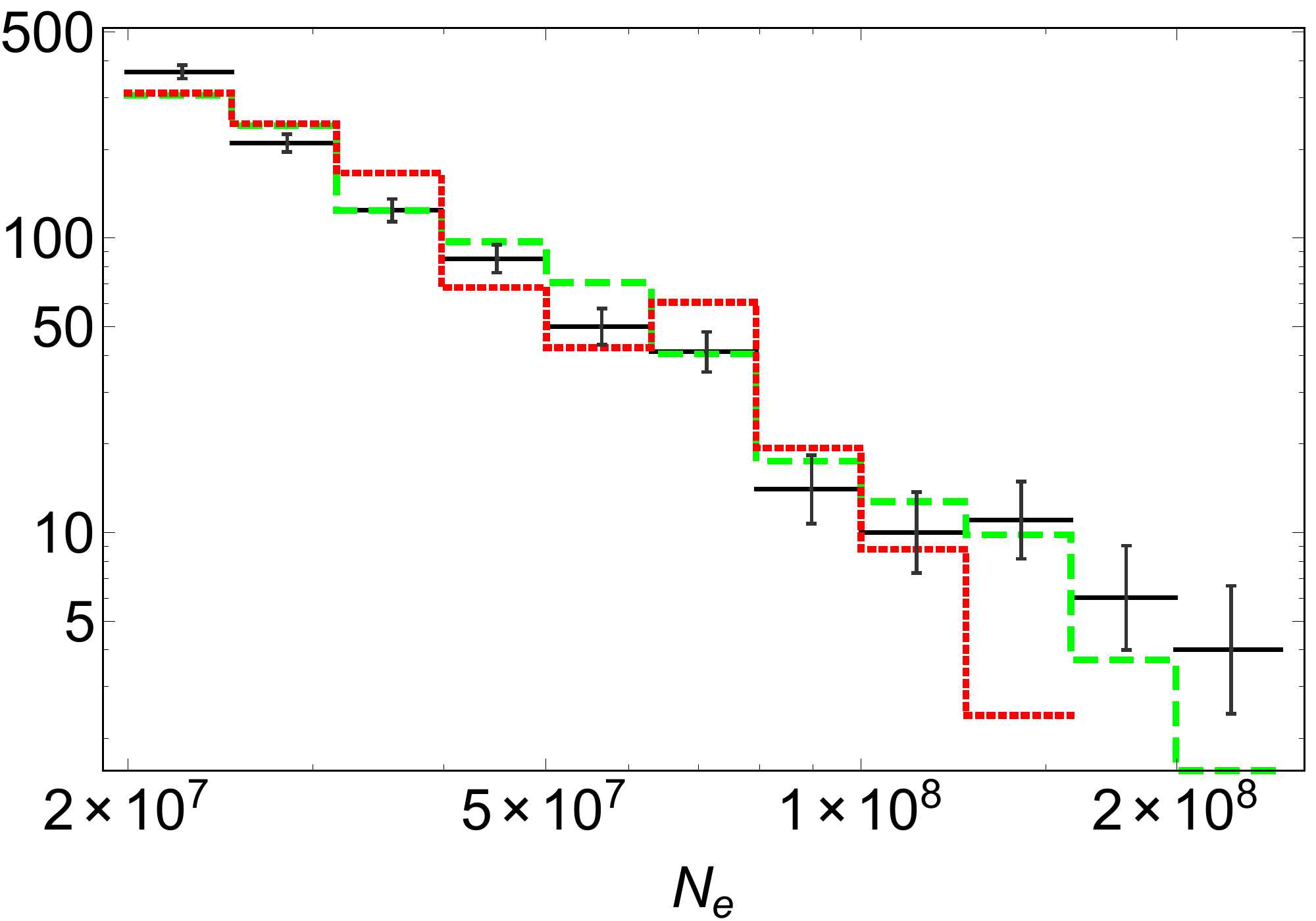} \\
(a)}
\end{minipage}
\hfill
\begin{minipage}[h]{0.48\linewidth}
\center{
\includegraphics[width=1\linewidth]{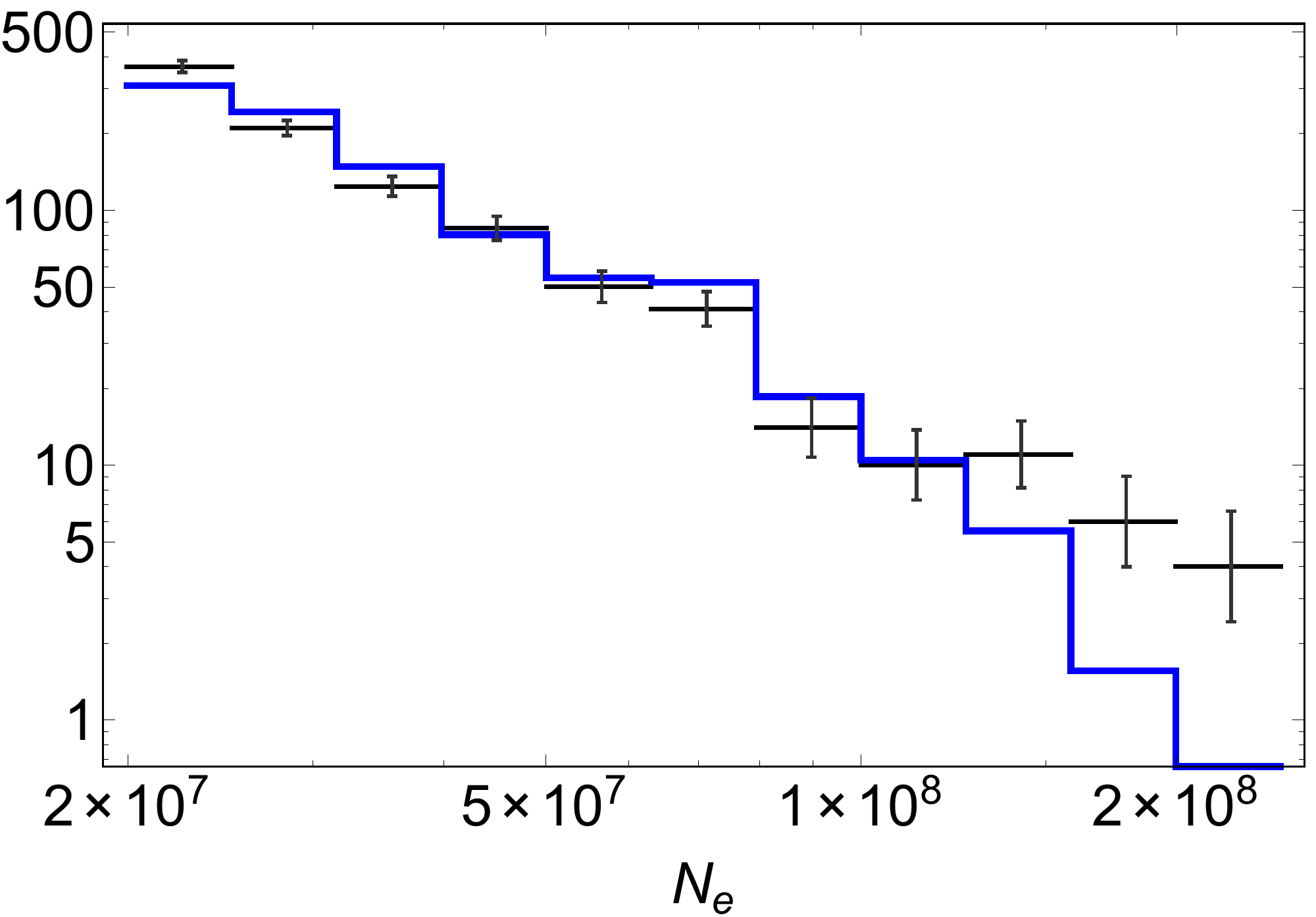}%
\\
(b)
}
\end{minipage}
\caption{Data and MC comparison of the distribution in $N_{e}$.
Points with error bars: data.
(a):~green dashed hystogram: MC (protons), red dotted hystogram: MC (iron);
(b):~blue hystogram: MC (the best-fit composition).
}
\label{fig:Ne-comparison}
\end{figure}

\paragraph{2. Shower size and the primary energy.}
The distribution in the reconstructed $N_{e}$ agrees well between the real
and simulated data sets, see Fig.~\ref{fig:Ne-comparison}.
This supports the use of the chosen thrown spectrum as a first
approximation.

\begin{figure}
\begin{minipage}[h]{0.48\linewidth}
\center{
\includegraphics[width=1\linewidth]{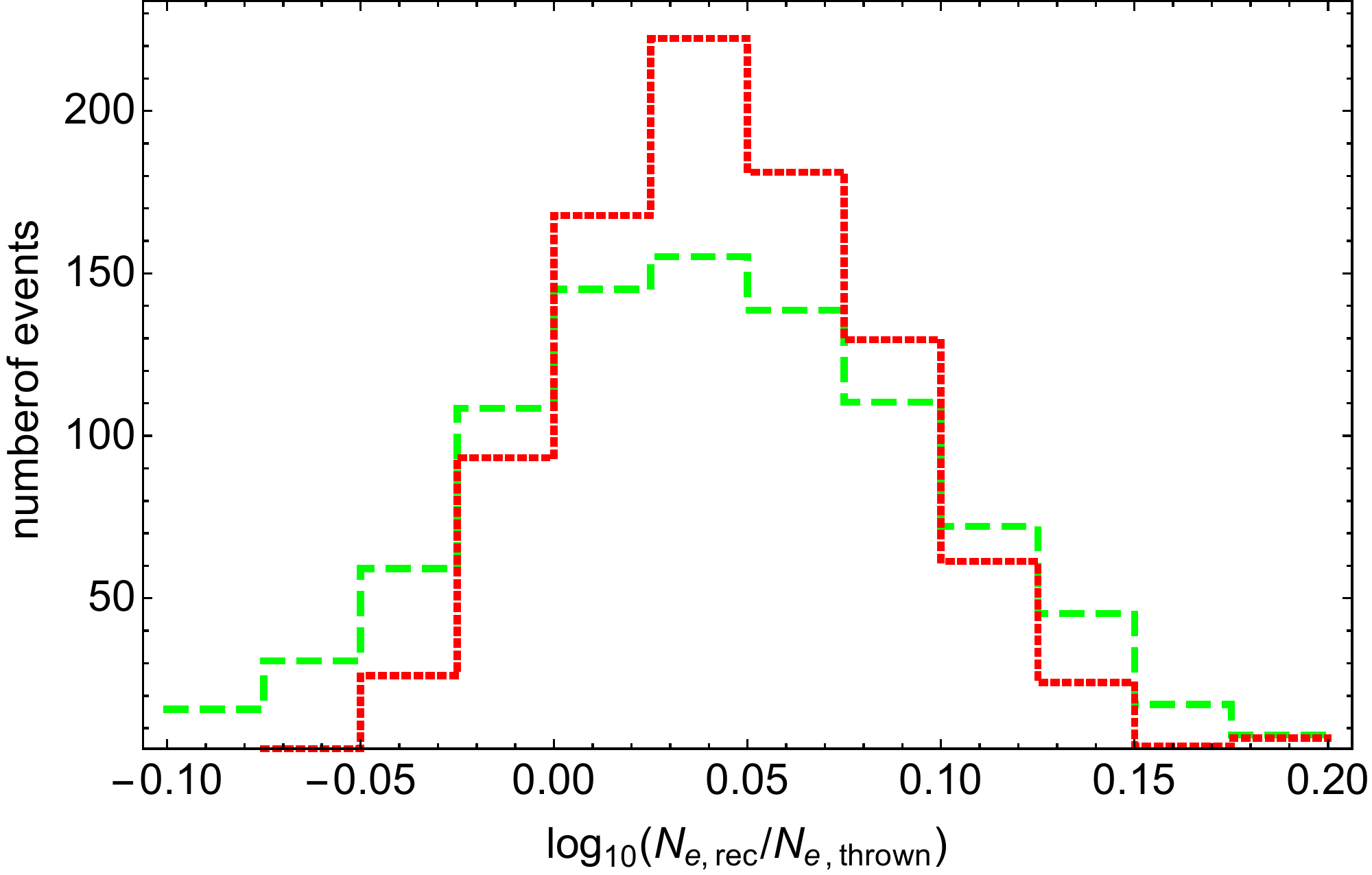} \\
\red (a) \black}
\end{minipage}
\hfill
\begin{minipage}[h]{0.48\linewidth}
\center{
\includegraphics[width=1\linewidth]{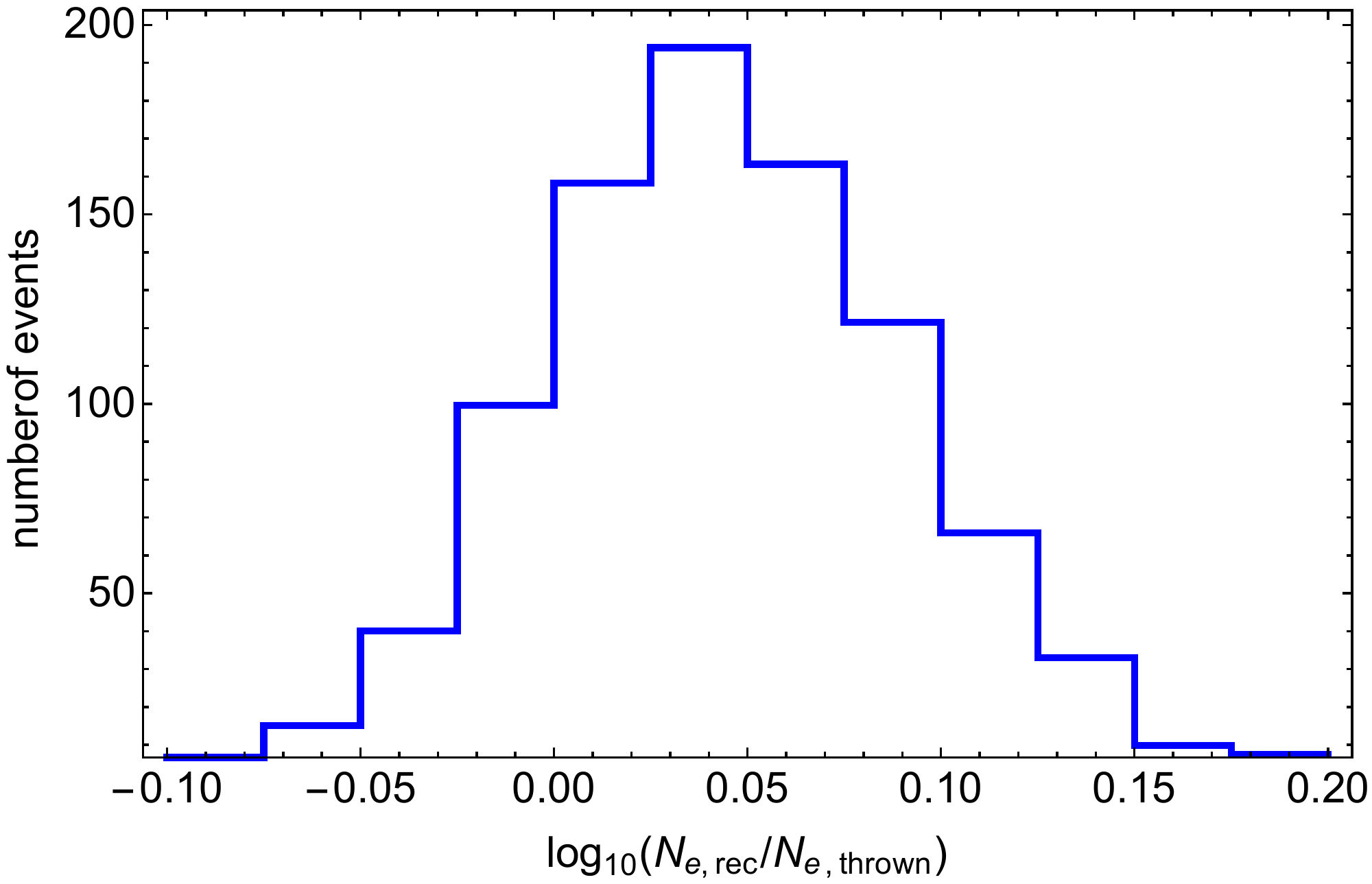}%
\\
(b)
}
\end{minipage}
\caption{Distribution of ratios of the reconstructed $N_{e}$ to the
``thrown'' $N_{e}$ which is the total number of charged particles from
the CORSIKA output supplemented by the contribution of photons to the
signal, see the text, for MC events.
\red (a):~green dashed hystogram: MC (protons), red dotted hystogram: MC
(iron); (b):~blue hystogram: MC (the best-fit composition). \black}
\label{fig:Ne-precision}
\end{figure}

Figure~\ref{fig:Ne-precision}
illustrates the accuracy of the reconstruction of $N_{e}$ in comparison
with the total number of charged particles and interacting photons in
counter calculated by CORSIKA. \red The relative bias of $\sim 7\%$ may be
attributed to the difference in the definitions of the number of charged
particles in a shower. \black The accuracy of the $N_{e}$ determination is
also presented in Table~\ref{tab:accuracy}. \red Note that it decreases
with energy reaching $\sim 0.15$ for the highest energies we consider.
\black

\begin{figure}
\center{
\includegraphics[width=\linewidth]{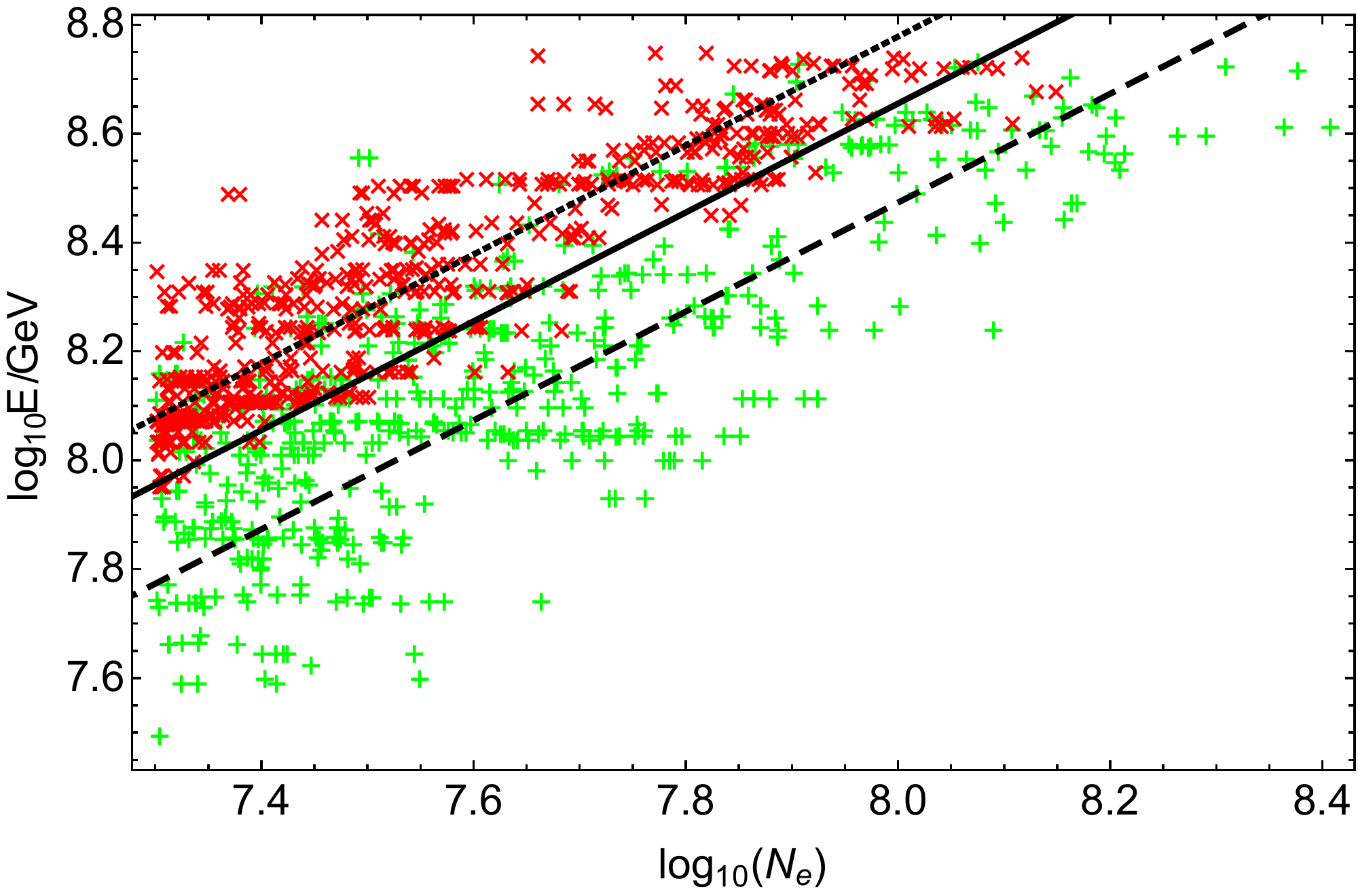}}
\caption{Thrown energy $E$ versus reconstructed $N_{e}$
for the MC
events: green pluses -- primary protons, red crosses --
primary iron. The full line represents the relation
(\protect\ref{Eq:E-Ne}), the dashed and dotted lines represent the
relations for primary protons and iron, respectively.
}
\label{fig:Ne-E}
\end{figure}

The relation between $N_{e}$ and the primary energy is model-dependent in
simulations and always depends on the type of the primary particle. From
our simulations, we may estimate it for the QGSJET-II-04 model we use.
Figure~\ref{fig:Ne-E}
presents, for the proton and iron primaries, the scatter plot of thrown
energies and reconstructed values of $N_{e}$.
The following relations may be obtained\red, through a fit,\black for
primary protons,
\begin{equation}
\log_{10}\left(\frac{E_{\rm p}}{\rm GeV}\right)=
(0.47 \pm 0.01) + \log_{10} N_{e}
,
\nonumber
\end{equation}
and for primary iron nuclei,
\begin{equation}
\log_{10}\left(\frac{E_{\rm Fe}}{\rm GeV}\right)=
(0.78 \pm 0.01) + \log_{10} N_{e}
,
\nonumber
\end{equation}
respectively. Here, we assumed that $E \propto N_{e}$. This is
justified for a short energy range we consider (one decade); allowing for
a power-law dependence, $E\propto N_{e}^{a}$, gives $a$ very close to
unity, but with a less stable fit. For practical
reasons, however, we need a ``$N_{e}$--$E$'' relation which may be used
without the knowledge of the primary type, as is relevant for the real
data. We make use of our working best-fit composition to obtain an average
relation,
\begin{equation}
\log_{10}\left(\frac{E}{\rm GeV}\right)=
0.65  + \log_{10} N_{e}
.
\label{Eq:E-Ne}
\end{equation}
Note that, contrary to similar relations for scintillator detector arrays,
Eq.~(\ref{Eq:E-Ne}) does not include the zenith angle $\theta$. This approximation is feasible
because the shower age is taken into account in the LDF and only
$\theta \le 30^{\circ}$ showers are analyzed.

\begin{figure}
\center{
\includegraphics[width=1\linewidth]{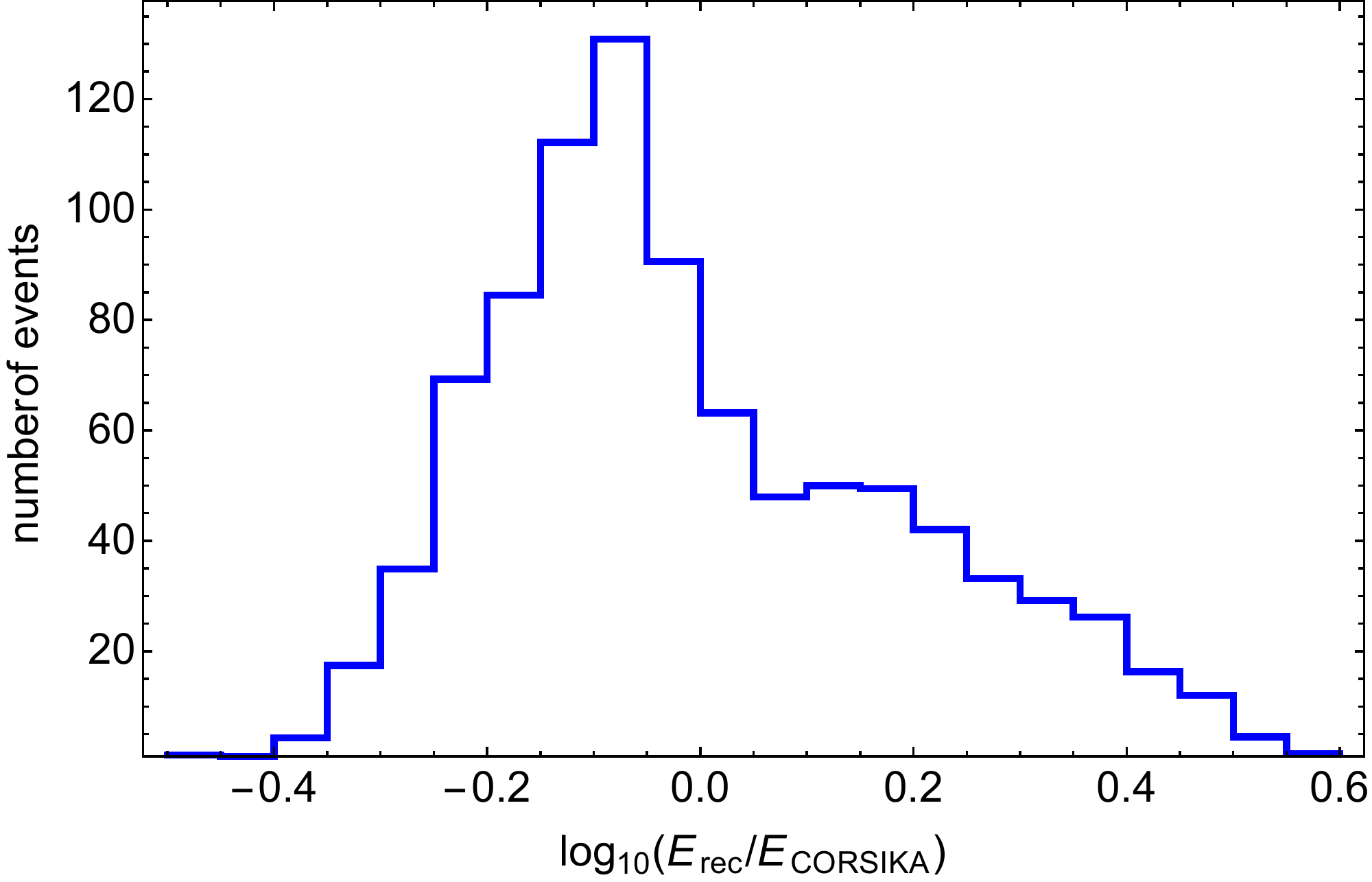}%
}
\caption{Distribution of ratios of the energy $E$, calculated by means
of Eq.~(\protect\ref{Eq:E-Ne}) from the reconstructed $N_{e}$, to the
thrown primary energy,
for the MC events (the best-fit composition).
}
\label{fig:E-precision}
\end{figure}

Having Eq.~(\ref{Eq:E-Ne}) as a working model for the energy estimation of
hadronic primaries, one may estimate the accuracy of the energy
reconstruction by comparison of the thrown MC energy and the energy
calculated from the reconstructed $N_{e}$ by means of Eq.~(\ref{Eq:E-Ne}).
This comparison is illustrated in Fig.~\ref{fig:E-precision}
and the resulting accuracy is quoted in Table~\ref{tab:accuracy}.

\begin{figure}
\begin{minipage}[h]{0.48\linewidth}
\center{
\includegraphics[width=1\linewidth]{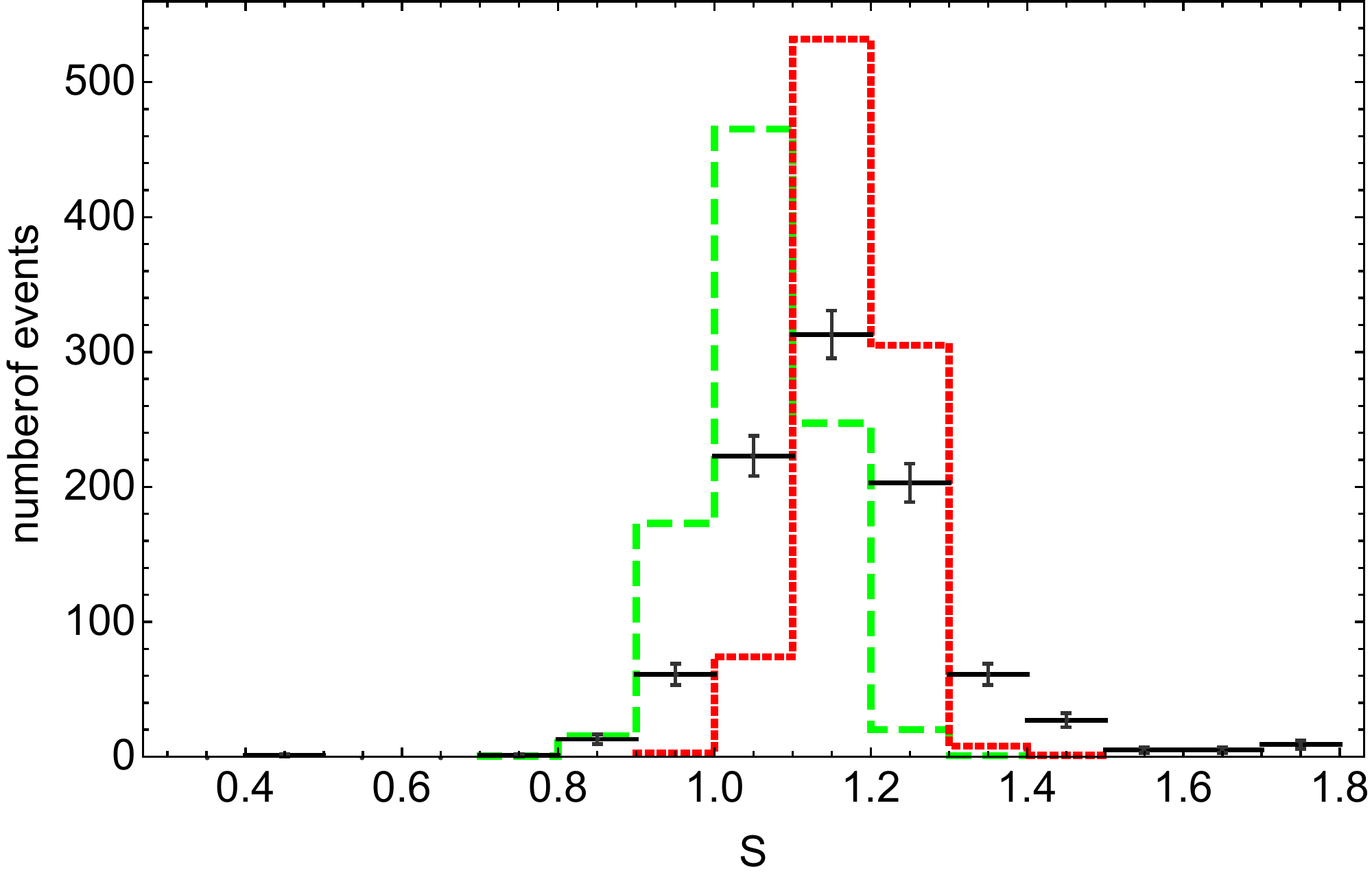}%
\\
(a)
}
\end{minipage}
\hfill
\begin{minipage}[h]{0.48\linewidth}
\center{
\includegraphics[width=1\linewidth]{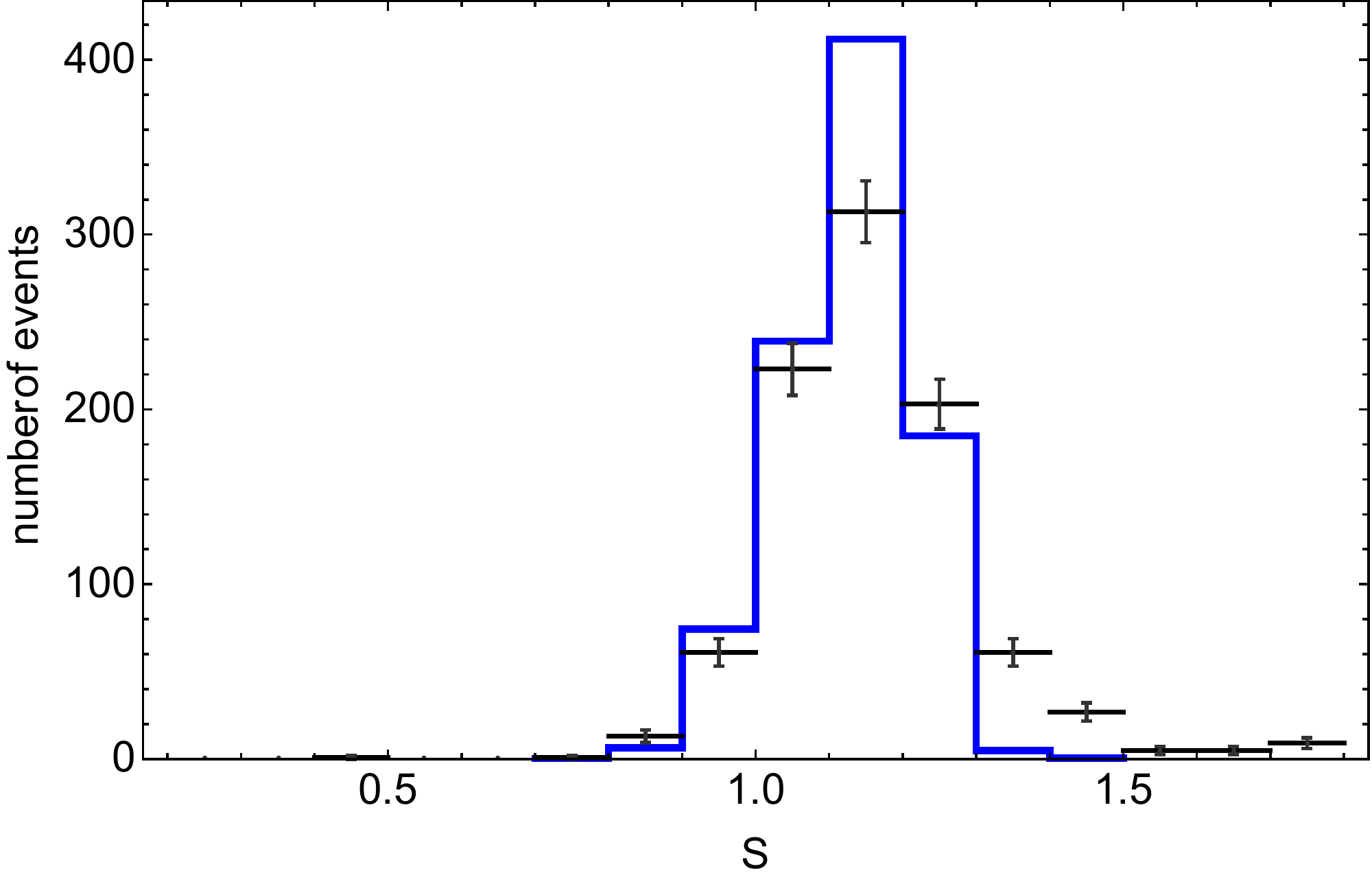}%
\\
(b)
}
\end{minipage}
\caption{
Data and MC comparison of the distribution in $S$. Points with error bars:
data. (a)
Green dashed hystogram: MC
(protons); red dotted hystogram: MC (iron).
(b) Blue hystogram: MC (the best-fit composition).
}
\label{fig:s-comparison}
\end{figure}

\paragraph{Shower age and the primary composition.}
The reconstruction of geometrical parameters discussed above is
insensitive to the type of the primary particle of an EAS.
However, the shower age parameter $S$, which is related to the depth of
the shower development in the atmosphere and is reconstructed in the LDF
fit, is composition-sensitive. As a result, the distributions of the MC
events in $S$ differ significantly for the proton and iron primaries, see
Fig.~\ref{fig:s-comparison}~(a).

\begin{figure}
\center{
\includegraphics[width=1\linewidth]{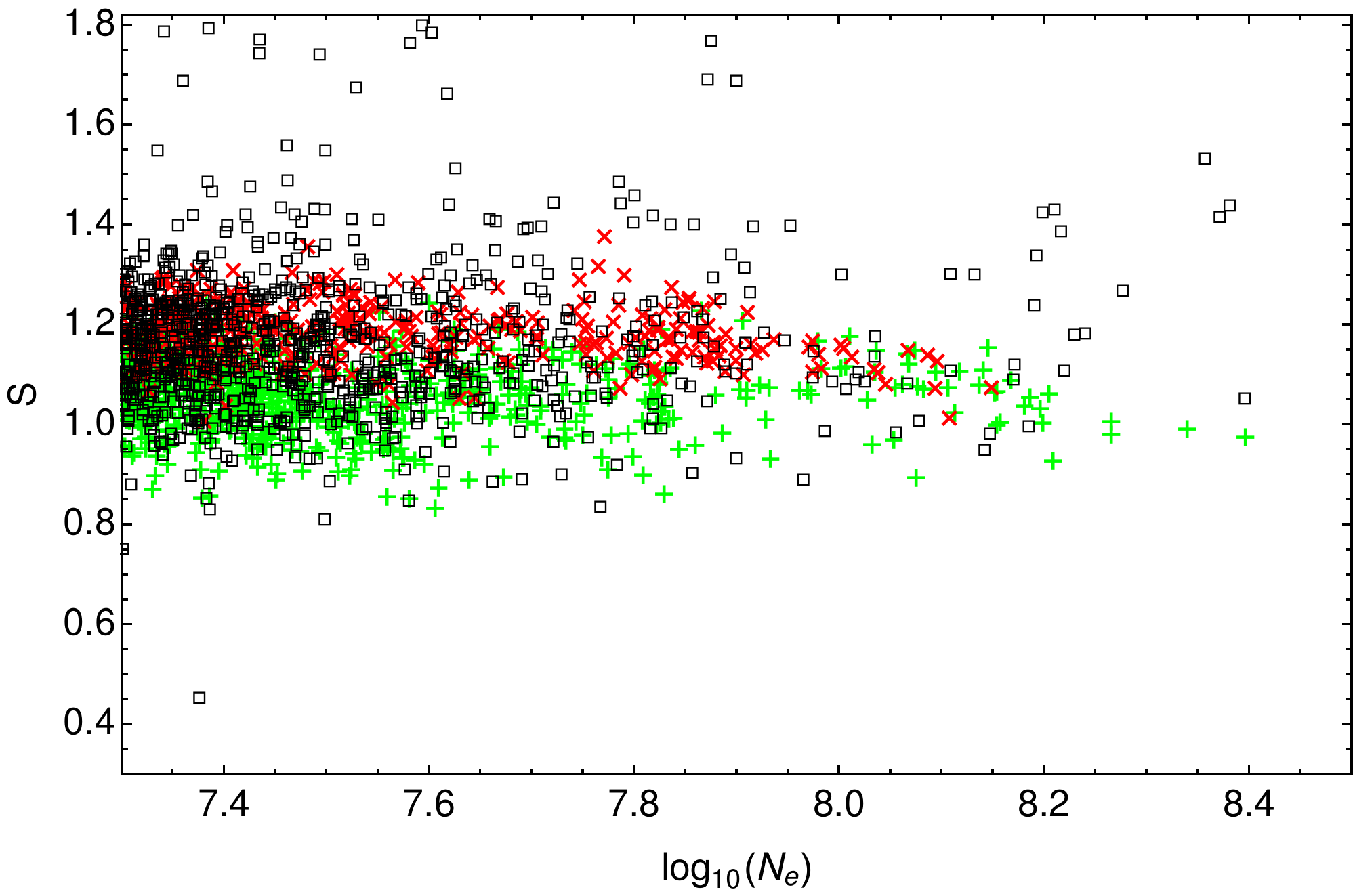}%
}
\caption{The shower age parameter
$S$ vs $N_{e}$. Boxes: data; green pluses: MC (protons); red
crosses: MC (iron).}
\label{fig:rhino}
\end{figure}

Therefore, we have used the data distribution over $S$ to
derive the relevant primary composition. The best-fit composition appears
to be almost energy-independent in the energy range we consider,
$\sim(10^{17}-10^{18})$~eV (though the highest-energy part of the data set
is statistically depleted). For the two-component model we use, it
includes $\approx 43\%$ of protons and $\approx 57\%$ of iron.
The comparison between data and MC mixed according to this
composition is presented in Figs.~\ref{fig:s-comparison}~(b),
\ref{fig:rhino}.
It is the composition we used for basic data/MC
comparison tests throughout the paper.
A good agreement of data and MC in $S$ obtained for the
realistic primary composition
represents a non-trivial
test of our full MC model of the EAS-MSU installation.
\red
The result obtained for the chemical composition from the fit of
the $S$ distribution with a proton/iron mixture may be considered as
approximate; it is the composition assumed for the considered model of the
detector only. Nevertheless our result shows good agreement with more
robust results of other experiments e.g.
KASCADE-Grande~\cite{KASCADE-comp} and Tunka-133~\cite{Tunka-comp}.

\paragraph{Discussion.}
The overall data/MC agreement is good, which justifies the use of our MC
model of the detector for future studies. However, a detailed examination
of the plots reveals minor disagreements in distributions of some
reconstructed observables, mainly in tails. In particular, there is \black
a minor excess of events with large $S$ in data with respect to
simulations. These events, about a few per cent of the data set, are
probably caused by some detector saturation effect related to occasional
unrecorded technical problems in the particle-detection system.
\red
Saturation effects can be caused by the technique of the
measurement of the charged-particle density, $\rho_{c}$. As it was
mentioned in Sec.~\ref{sec:array}, each detector station in the
peripheral system contains 72 large counters, 24 medium counters and 24
small counters. The charged particle density is estimated as $\rho_c = \ln
\left(n/(n-m)\right)/A $, where $n$ is the number of counters in the
detector station, $m$ is the number of fired counters and $A$ is the area
of the counter.
This yields the following limits of $\rho_c$ measurement:
from $0.42$~m$^{-2}$ to $130$~m$^{-2}$ for large counters,
from $4.3$~m$^{-2}$ to $318$~m$^{-2}$ for medium counters
and from $23.6$~m$^{-2}$ to $1766$~m$^{-2}$ for small counters.
As a default, for every detector station, $\rho_c$ is
calculated from readings of the large counters. However, once
all the large counters are fired, $\rho_c$ is calculated from readings of
medium counters. The same rule applies to medium and small counters.
There could exist, however, situation when all large or medium counters are
hit by shower particles, while some of the counters didn't trigger due
to the hardware failure. This could lead to a significant underestimation of
$\rho_c$ at this detector station. \black This effect can only reduce the
particle density recorded by a given detector close to the shower axis,
leading to a flatter LDF (i.e. to higher $S$ and $N_e$ parameters) \red
and to a light overestimation of $R$. \blue The latter \red happens due to
high concentration of detector stations in the center of array. Actually,
the possibility of one or two undetection errors in each detector station
is taken into account in the reconstruction program but other errors of
the same nature still could lead to the same effect, which we call the
saturation effect. While the usual saturation is accounted for in all our
simulations, we are unable to trace these possible undocumented errors and
to model them. We suppose that these unresolved errors \black are
responsible for the minor discrepancy between MC and data for small $R$ in
Fig.~\ref{fig:R-comparison}, for large $N_e$ in
Fig.~\ref{fig:Ne-comparison} \red (and, consequently, for large $E$ in
Fig.~\ref{fig:E-precision}) and for large $S$ in
Figs.~\ref{fig:s-comparison}, \ref{fig:rhino}. \black Note that these
events imitate old showers and cannot mimic gamma-ray induced events,
which develop deeper in the atmosphere, see e.g.\ Refs.~\cite{HomolaRiess,
Nuhuil}, while the study of primary gamma rays, to be reported in a
separate publication, is the main motivation for the present work.

\section{Conclusions}
\label{sec:concl}
To summarize, we presented a full chain of Monte-Carlo simulations of air
showers developing in the atmosphere, being detected by the EAS-MSU array
and reconstructed by the procedure equivalent to that used for the real
data. We have verified that the simulation describes well the distribution
of the data in basic parameters. Assuming a modern hadronic-interaction
model, we obtained a relation between the reconstructed shower size
$N_{e}$ and the primary energy $E$. We used our Monte-Carlo event sets to
estimate the accuracy of the reconstruction of the shower geometry,
$N_{e}$ and energy. Forthcoming studies will use these simulations to
reanalyze data on $E_{\mu}>10$~GeV muons in the air showers
recorded by the installation, which will open the possibility to test,
within modern frameworks, the origin of the apparent excess of muonless
events seen in the EAS-MSU data.

\acknowledgments

The experimental work of the EAS-MSU group
was supported in part by the grants of the Government of the
Russian Federation (agreement 14.B25.31.0010) and of the Russian
Foundation for Basic Research (project 14-02-00372).
Development of the methods to search for primary photons and application
of these methods to the EAS-MSU data were
supported by the Russian Science Foundation, grant 14-12-01340.
\red Numerical simulations have been performed at the computer cluster of
the Theoretical Physics Department of the Institute for Nuclear Research
of the Russian Academy of Sciences. \black


\end{document}